\documentclass[11pt,letterpaper]{article}
\usepackage[margin=1in]{geometry}
\usepackage{amsthm}
\newtheorem{theorem}{Theorem}
\newtheorem{lemma}{Lemma}
\newtheorem{claim}{Claim}
\theoremstyle{definition}
\newtheorem{definition}{Definition}
\date{}
\usepackage[T1]{fontenc}
\usepackage{amsmath}
\usepackage{amssymb}
\usepackage{wrapfig}
\usepackage{comment}
\usepackage{subcaption}
\usepackage{tikz} 
\usepackage{epigraph}
\newcommand{\eps}{\epsilon}
\usepackage[hidelinks]{hyperref}
\usepackage{cleveref}
\usepackage{mdframed}

\usepackage{algorithm}
\usepackage{algpseudocode}

\usepackage{newtxtext}       %
\usepackage[varvw]{newtxmath}       

\usepackage{graphicx}
\begin{document}

\title{An Efficient Algorithm for Matroid Cups}

\author{Stephen Arndt\thanks{Carnegie Mellon University, Tepper School of Business. Supported in part by NSF grant CCF-2550399.} \and Benjamin Moseley\thanks{Carnegie Mellon University, Tepper School of Business. Supported in part by NSF grant CCF-2550399. } \and Kirk Pruhs\thanks{University of Pittsburgh. Supported in part by NSF grants CCF-2550399 and CCF-2209654.} }

\maketitle              
\begin{abstract}
We design a polynomial-time algorithm for matroid  cups with $O(1)$ discrepancy. 
Our matroid cups algorithm design consists of a sequence of reductions 
to simpler instances, culminating in $O(n)$ instances of matroid cups where the matroid has rank 1.
\end{abstract}

\section{Introduction}
We consider a setting consisting of a set system $M=(U, {\mathcal I})$ and a vector $x$ of probabilities. 
The ground set $U$ is a collection of $n$ operations/actions that one wants to perform periodically,
and for each action $u \in U$ the $u^{th}$ coordinate of $x$, $x_u$, represents the desired frequency for
action $u$. 
A subset $B$ of $U$ is independent in $M$, that is in $\mathcal I$, if and only if all the actions in $B$ can be performed simultaneously. 
A schedule $S$ then specifies for each positive integer time $t$ a subset $S_t$ of $U$ with the property
that $S_t \in {\mathcal I}$. 
The height of  $u$ at time $t$, $h_u(t)$, is $x_u (t-s)$, where $s$ is the latest
time prior (or equal to) $t$ that action $u$ was taken. 
The discrepancy $\Delta_u(t)$ of  $u$ at time $t$ is the
absolute value of the difference of the number of times the action has been performed by time $t$ 
and $x_u t$. 
In the ``bamboo'' problem  the objective is to minimize the maximum height, that is minimize:
$$\max_{t}   \max_{ u \in U}  h_u(t)$$
In the ``cups'' problem the objective is to minimize the maximum discrepancy, that is minimize:
$$\max_{t}   \max_{ u \in U}  \Delta_u(t)$$
We investigate whether there is an efficient algorithm, 
that given  an independence oracle for $M$, will iteratively output the components $S_1, S_2, \ldots $ of an infinite schedule $S$, where by efficient we mean  that the time used by the algorithm to output each component
is bounded by a polynomial in the cardinality $n$ of the ground set $U$ and the number of bits
used to present the probability vector $x$. Before stating the results of our investigation, 
we give some background from the literature.

\subsection{The Story to Date}

An algorithm/schedule that achieves discrepancy $d$, also achieves height $2d$, but the converse does not hold
as there are instances where height $O(1)$ is achievable but discrepancy $O(1)$ is not achievable~\cite{Mendoza-CadenaM26}.  So low discrepancy is harder to achieve than low height. There are simple 
instances for which every schedule has discrepancy at least  $1-\epsilon$ and height at least $2 - \epsilon$,
for example if $x_u = 1 - \epsilon$, $x_v = \epsilon$ and only one of the actions $u$ or $v$ can be performed
at each time. So the gold standard for discrepancy is 1 and the gold standard for height is 2.

For all set systems, there exist schedules with maximum height $O(\log n)$, and there
are set systems where this is best possible~\cite{cicerone2019,Mendoza-CadenaM26}.
The $\Omega( \log n)$ lower bound for general set systems
motivates the consideration of set systems that are matroids.
For every matroid $M$, there exists a schedule with optimal discrepancy 1~\cite{Mendoza-CadenaM26},
but it was not know how to compute an $O(1)$-height or $O(1)$-discrepancy for a general matroid  in polynomial-time; In this setting a polynomial-time algorithm is one where for all times $t$ the amount of time the algorithm runs between producing $S_{t-1}$ and $S_t$ is polynomial
in $n$, the cardinality of the ground set.

However polynomial-time algorithms are known for special classes of matroids. 
Most importantly 
there exists 
a polynomial-time algorithm for rank 1 matroids that achieves optimal discrepancy 1 (and hence optimal height 2)~\cite{tijdeman1980}.
We will conceptualize this algorithm, which we  will refer to as the Unit Clock  Algorithm,  as a circular clock,
 with one hand for each $u \in U$. All the hands initially have the same initial orientation. 
Then at each integer time,  each hand  $u \in U$ advances clockwise $2 \pi x_u$ radians,
and the clock outputs the $u $ that will soonest reach the original orientation,
among those $u$ that have not been output since they last passed the original orientation. 
Further there  are efficient algorithms that achieve discrepancy $2$ for uniform matroids~\cite{bender2019},  
height 2 for uniform and partition matroids~\cite{Mendoza-CadenaM26},
and height 4 for laminar and graphic matroids~\cite{Mendoza-CadenaM26}.

These are the most relevant results for our purposes, but this just scratches the surface on research in this area
originating from  the discrete math and combinatorial optimization communities. Good jumping points for further background
are the recent papers \cite{Mendoza-CadenaM26,kuszmaul2022}. Additionally there is a whole research community, the real-time scheduling community, devoted to studying periodic scheduling. A reasonable jumping off point for research from this community is the textbook \cite{baruah2015multiprocessor}.

\subsection{Our Contribution to the Story}

We first note that the existential result from \cite{Mendoza-CadenaM26} can relatively easily be made constructive
to yield an efficient algorithm for matroid bamboo for that achieves
essentially optimal height for all matroids.

\begin{theorem} \label{thm:matroidbamboo}
 For all fixed $\epsilon > 0$ there is a polynomial-time algorithm for matroid bamboo that achieves height at most $2 + \epsilon$. 
\end{theorem} 

The proof of Theorem~\ref{thm:matroidbamboo} appears in Appendix~\ref{app:matroidbamboo}. This algorithm for matroid bamboo does not necessarily yield schedules with $O(1)$ discrepancy. 
So our  main contribution  is an efficient algorithm for matroid cups for that achieves
$O(1)$ discrepancy for all matroids. 

\begin{theorem}
    \label{thm:matroidcups}
There is a polynomial-time algorithm for matroid cups that achieves discrepancy at most $33$.
\end{theorem}

In the next section we give an overview of the design and analysis of the algorithm 
from Theorem \ref{thm:matroidcups}, and the organization of the rest of the paper.

\section{Technical Overview}

Our algorithm design for matroid cups consists of a sequence of reductions 
to simpler instances of matroid cups, culminating in $O(n)$ instances of matroid cups where the matroid has rank 1, thus solvable by the  Unit Clock algorithm. To aid the reader
in seeing the main ideas, we brush some details under the rug in this section.

Our first reduction reduces an instance of matroid cups for the original general matroid $M$ to an instance of matroid cups
for a  $M_0$ that is the contraction of a transversal matroid that can be represented by a polynomially-sized bipartite graph. 
To accomplish this reduction, our algorithm creates its  schedule $S$ in blocks of $H$ consecutive units of time.
The schedule  $S_t$ for block $t$  is computed at the start of time $(t-1)H$ +1,
and will be executed from time $(t-1)H +1$ to $t H$. 
The services during $S_t$ will consist of a multiset $Z_t$ of $r H$ elements of the ground set $U$ of $M$, where $r$ is the rank of $M$. The multiset $Z_t$ will be the union of two multisets, the required services 
$R$ that will appear in every $Z_t$, and the fill $I_t$. 
The required services  $R$ can be represented by a vector $\ell$ where  $\ell_u = \lfloor x_u H \rfloor$  is
the number of required services for
each $u \in U$.  
 The actual schedule $S_t$ is then produced from $Z_t$ by 
a  matroid intersection algorithm  in a standard manner. 
This reduces the matroid cups problem for the original matroid $M$ to 
the problem of producing the sequence $I_1, \ldots $ of fills.
But the problem of producing the sequence of fills is just an instance 
of the cups problem for the set system $M_0  = (U, \mathcal{I}_0)$,
where a subset $I \subseteq U$ is independent (that is $I \in {\mathcal I}_0$) if 
 $R  \cup I$ can be feasibly scheduled in $H$ time units,
and where the probability associated with each element $u \in U$
is $y_u = x_u H - \ell_u$.
A key insight is that if $H \geq n^2$ then $M_0$ is equal to a matroid $M_1$, which is the contraction on $R$ of
a transversal matroid on ground set $R \cup U$.
Further, this transversal matroid is represented by a polynomial-size bipartite graph $G = ( R \cup U, Z, E)$, where the resource side
$Z$ has cardinality $r H$. We give further detail in Section \ref{sect:transversal}.

Our next reduction step is to reduce the matroid cups problem for matroid $M_0$ to the 
 matroid cups problem for an even simpler matroid $M_2 = ( U , {\mathcal I}_2)$,
 which is a contraction of the transversal matroid represented by the forest $G_2 = ( R \cup U, Z, F)$,
 derived
 from $G$ by replacing the edge set $E$ with an arbitrary acyclic fractional matching $F$ in $G$ 
that covers each element of $R$ to extent 1, each element of $Z$ to extent 1,  and each $u \in U$ to extent $y_u $.
 The matroid $M_2$ is also a series-parallel graphic matroid 
represented by a polynomial-size series-parallel forest $T$.~\footnote{This observation is not completely novel as Speyer and Williams \cite[Section 7.2, Remark 7.5]{SpeyerWilliams2021} remark that the principal transversal matroid associated with a bipartite tree $F$ is series-parallel, referencing \cite[Proposition 5.1]{Speyer2008} for a hint at the proof, but do not give an explicit construction. }
The forest $T$ can be obtained from  $F$
 by rooting each tree in $F$ at an arbitrary element of
$Z$,  making nodes in $Z$ parallel nodes, and nodes in $R \cup U$ series nodes. The leaves of $T$ will be made to be the ground set $U$.
Thus the matroid $M_2$ is a reduction of the matroid $M_0$ in that every set that is independent in $M_2$ is also independent in $M_0$, and each fill $I_t$ returned by our algorithm will be a base of $M_2$.

We then observe that there is a simple randomized process $AR$ that generates a basis $I$ of matroid $M_2$ that matches the target marginal
probabilities in $y$ exactly, that is  for all $u \in U$ the probability
that $u \in I$ is $y_u$. 
In this process  each internal node $v \in T$ receives a bit from its parent (other than the root which always receives a 1 bit) and passes along bits to its children using a fixed probability distribution. 
For a parallel node $v$ with child $w$ define $\alpha_w$, to be the probability that $v$ passes  a received 1 bit
to $w$, and for a sequential node $v$ with child $w$, define $\alpha_w$ to be the probability that $v$ passes  a received 0 bit
to $w$. 
The basis $I$ is then the leaves that receive a 1 bit. We give further detail about the reduction to the series-parallel matroid $M_2$ and
the $AR$ process in Section \ref{sect:seriesparallel}.

\begin{figure}[htbp]
\centering
\begin{subfigure}{0.48\textwidth}
  \centering
  \includegraphics[width=\linewidth]{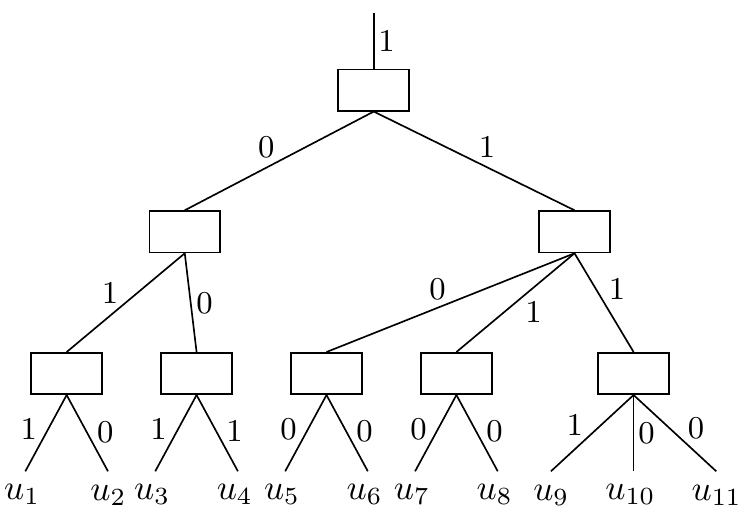}
  \caption{A chunk tree, in which the chunks are the rectangles, showing a computation that would output $I_t = \{u_1, u_3, u_4, u_9\}$.}
  \label{fig: ChunkTree}
\end{subfigure}
\hfill
\begin{subfigure}{0.48\textwidth}
  \centering
  \includegraphics[width=\linewidth]{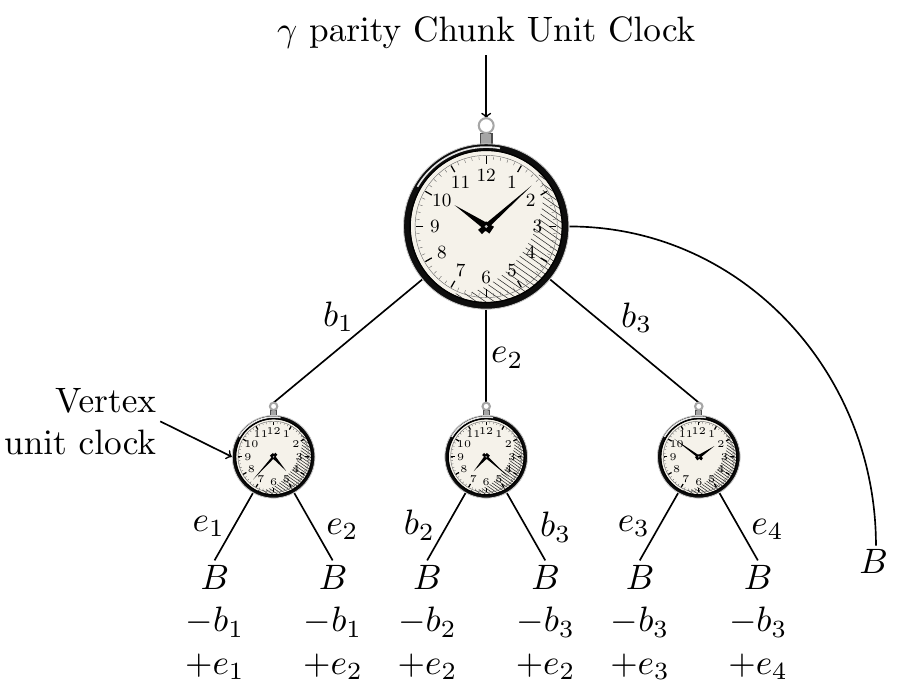}
  \caption{$\gamma$ parity half of  a chunk algorithm. The default basis is $B = \{b_1, b_2, b_3\}$. }
  \label{fig:ChunkInternalClock}
\end{subfigure}
\caption{Overview of our  algorithm $A$  for producing the fill.}
\label{fig:both}
\end{figure}

Thus our goal becomes designing
a deterministic procedure $A$, that produces an infinite sequence $I_1, \ldots$ of bases of $M_2$ that will constitute the fills,
and 
that maintains small discrepancy relative the $AR$ process.
To accomplish this, we  partition the internal nodes in $T$ into 
parts, that we call chunks. In our algorithm $A$ for  generating the next fill, 
each chunk will receive a bit from its parent chunk, and will pass along a bit to
each child chunk. See Figure \ref{fig: ChunkTree}. Note the problem of matching the $AR$ process internal to a chunk $C$ is 
just an instance of matroid cups where the ground set is essentially the edges leaving the chunk. 
We have two goals for when forming  the chunks: 
\begin{itemize}
    \item The chunks are sufficiently simple that we can design
an efficient  low-discrepancy algorithm for these chunk instances of matroid cups. 
\item The chunks significantly dissipate/distribute the
discrepancy received from above. 
\end{itemize} 
To better understand this second goal consider a parallel node $v \in T$, with two children $w_1$ and $w_2$. Assume that in the $AR$ process, the node $v$ sees 
a stream of all 1 bits from its parent, and in response to each bit  passes the $1$ to 
$w_1$ with probability $\alpha_{w_1}$ and  to 
$w_2$ with probability $\alpha_{w_2}$ (and passes the other child a 0 bit).
Suppose the discrepancy introduced above by our algorithm $A$ above on the incoming stream at $v$ is $\Delta$.
So $v$ sees $\Delta$ 0 bits.
If $\alpha_{w_1}=1$ and $\alpha_{w_2}=0$, then 
the discrepancy of the bits arriving at $w_1$ would be $\Delta$, plus any additional discrepancy introduced by
our algorithm at $v$. In contrast, if $\alpha_{w_1}=\frac{1}{2}$ and $\alpha_{w_2}=\frac{1}{2}$, then 
discrepancy of the bits arriving at both $w_1$ and $w_2$ would be $\Delta/2$, plus any additional discrepancy introduced by
our algorithm at $v$. Thus we want to form the chunks so that the effect of discrepancy introduced higher in $T$ by our algorithm will decrease geometrically as the bits flow toward the leaves.

To accomplish these two chunk formation goals, each chunk $C = v_1, \ldots, v_k$  will be ancestor-to-descendant paths,  
where each $v_i$ is the favorite child of $v_{i-1}$, and $\prod_{i=1}^k \alpha_k \approx \frac{2}{3}$,
where the favorite child is the one with maximum $\alpha$ probability. We give more detail of how the
algorithm chunks $T$, and uses the chunks to create the fill, in Section \ref{sect:chunkingT}.

After forming the chunks, we are left with the problem of designing an efficient 
chunk algorithm, which is just an algorithm for instances of matroid cups that arise in the chunks. 
Consider a chunk $C=v_1, \ldots, v_h$ which is an ancestor-to-descendant path in $T$. 
Let the children of nodes in $C$ (that are not in $C$) be  $L(C)=w_1, \ldots, w_{h+1}$. 
Thus our chunk algorithm determines the nodes in $L(C)$ that receive a 1 bit
in response to the node $v_1$ receiving a bit  $\gamma$, which we will call a parity bit. 
For each $\gamma \in \{0, 1 \}$ the algorithm maintains a default subset/base $B^\gamma$ of $L(C)$. 
Further for each $\gamma \in \{0, 1 \}$ the algorithm maintains a unit clock on the vertices in $L(C)$. 
In this context ``unit clock'' means a black box solution to a particular instance
of matroid cups with a matroid of rank 1 with discrepancy 1. 
These two clocks are called the chunk clocks. 
Further for each $\gamma \in \{0, 1 \}$, and for each vertex   $w \in   L(C)$, the algorithm maintains a unit clock for $w$,
which we call a vertex clock.
When the parity bit $\gamma$ arrives, the $\gamma$ parity chunk clock is called, producing an element $w_i \in L(C)$ (or possibly no element).
Then the $\gamma$ parity vertex clock for vertex $w_i$ is called, producing another element $w_j \in L(C)$.
If $w_i \in B^\gamma$ then the vertices in $L(C)$ receiving a 1 bit are those in $B^\gamma - w_i + w_j$,
and otherwise it will be the case that $w_j \in B^\gamma$, and then the vertices in $L(C)$ receiving a 1 bit are those in $B^\gamma - w_j + w_i$. So for each parity bit $\gamma$, the elements of $L(C)$ receiving a 1 bit are within one swap of a default base $B^\gamma$. Note that time only advances for a clock when it is called. See Figure \ref{fig:ChunkInternalClock} for an illustration of our chunk algorithm. We give more detail about the design
of the chunk algorithm in Section \ref{sect:chunkalgorithm}.

Finally in Section \ref{sect:analysis} we prove that our algorithm has $O(1)$ discrepancy. 
At a high level this follows from the fact that the chunk algorithm has $O(1)$ discrepancy 
for the matroid cups instances that arise internally to a chunk, and the fact that the 
effect of the discrepancy introduced within a chunk decreases geometrically as the bits flow down 
the chunk tree.

To close our overview let us recall Segal's law~\cite{quoteinvestigator2022watch}:
\begin{quote}
    ``A man with one watch knows what time it is. A man with two watches is never sure.''
\end{quote}
In fact this quote is misattributed to Segal, and the originator of the quote remains unknown~\cite{quoteinvestigator2022watch}.  So this allows us to irresponsibly   conjecture that this quote originated from some early researcher 
 perplexed by  the puzzle of matroid cups. In the implausible circumstance that this conjecture is correct, 
one can view our algorithm  as overcoming the challenges expressed in Segal's law,
by deriving a good time for each action from the state of many clocks. Our top priority  was to explain how to achieve $O(1)$ discrepancy in polynomial time as cleanly as possible; in particular, we did not prioritize obtaining the best possible constants.

\section{Reduction To A Contraction of a Transversal Matroid }
\label{sect:transversal}

We now want to understand the subsets of the ground set $U$ that can feasibly be chosen as the fill. We start with the characterization used in ~\cite{Mendoza-CadenaM26}, and then introduce some definitions in preparation for giving our alternative characterization.

\begin{lemma}\cite{Mendoza-CadenaM26,schrijver_book}
    A subset $I \subseteq U$ can be feasibly scheduled with the services in $R$ in $H$ time units  if and only if  $\ell+ \mathbf{1}_I  \in H \cdot P(M)$, where $\mathbf{1}_I$ is the indicator vector for $I$, and $H \cdot P(M)$ is the independence polytope for matroid $M$  inflated by a factor of $H$.
\end{lemma}

\begin{definition} \label{defn:lambda}
    Let  $x = \sum_{i=1}^n \lambda_i \mathbf{1}_{B_i}$ for bases $B_i$ of $M$, $i \in [n]$ where $\sum_{i=1}^n \lambda_i = 1$ and $\lambda_i \geq 0$ for all $i \in [n]$. Assume without loss of generality that $\lambda_1 \ge \lambda_i $ for $i \in [2, n]$.
    For convenience, let $B = B_1$. 
\end{definition}

\begin{definition}[Fill Set System]\label{def:fill_set_system}
Let $H \geq 1$ be a positive integer. The fill set system $M_0(H) = (U, \mathcal{I}_0)$ is the set system where $I \subseteq U$ is in $\mathcal{I}_0$ if and only if $\ell+ \mathbf{1}_I \in H \cdot P(M)$. 
\end{definition}

\begin{definition}
 We define $G(H) = (U \cup R, Z, E)$ as the bipartite graph where
\begin{itemize}
     \item The left-hand side $U \cup R$ consists of $\ell_u+1$ vertices for each element $u \in U$. 
     \item The right-hand side $Z$ consists of $H$ vertices for each element $b \in B$. 
     \item For each element $b \in B$ the $\ell_b+1$ left-hand side vertices  that are derived from $b$ are connected 
     by an edge to the $H$ right-hand side vertices derived from $b$.
     \item For each element $u \in U - B$ the $\ell_u+1$ left-hand side vertices  that are derived from $u$ are connected 
    by an edge to each right-hand side vertex derived from an element $b \in B$ for which $B - b + u $ is independent in the original matroid $M$.
 \end{itemize}
\end{definition}

\begin{definition}
    Define the set system $M_1(H) = (U, {\mathcal I}_1)$ to be the one where $I \subseteq U$ is in ${\mathcal I}_1$ if and only if there exists a matching in $G(H)$ that matches every
    element in $I \cup R$.
\end{definition}

\begin{lemma}\label{lemma:M1Matroid}
    $M_1(H)$ is a matroid.
\end{lemma}

\begin{proof}
$G(H)$ represents  a transversal matroid  with ground set $U \cup R$, and $M_1(H)$ 
results from contracting this transversal matroid on $R$.
Thus $M_1(H)$ is a matroid as matroids are closed under contraction.
 \end{proof}

\begin{lemma}\label{lemma:qh_in_sh}
For all $H \ge n^2$, every independent set of $M_1(H)$ is an independent set of $M_0(H)$.
\end{lemma}

\begin{proof}
Let $I \in \mathcal{I}_1$ be arbitrary. By the definition of $M_1(H)$ there
exists a matching $\Gamma$ in $G(H)$ that matches every element in $I \cup R$. We then want to show $I $
is independent in $M_0(H)$. By the definition of $M_0(H)$, we need to show that  $\ell+ \mathbf{1}_I \in H \cdot P(M)$. 
A point $z \in P(M)$  if and only if for all subsets $S \subseteq U$ it is the case that $\sum_{u \in S} z_u \le r(S)$, where $r$ is the rank function for $M$~\cite{schrijver_book}.
Thus is sufficient to show that for all $S \subseteq U$ it is the case that
\begin{equation}
    \sum_{u \in S} \ell_u  + |I \cap S| \le H \cdot r(S)    \label{eqn:qhinsh-proof}
\end{equation}
So consider an arbitrary such $S$. Let  $T \subseteq U \cup R$ be the collection of all the
elements in $R$ that are derived from some element in  $S$ union with all elements in $I \cap S$.  Let $N(T)$ be the neighborhood of $T$ on the right-hand side of  $G(H)$. At a high level, the insight behind the proof is that if the neighborhood
of $T$  in $G(H)$ is small, then the matching in $G(H)$ guarantees that the matroid constraint for $S$ holds,
and if the neighborhood of $T$ in $G(H)$ is large, then there must be some slack in the matroid constraint for $S$, and thus making $H$ large enough handles rounding the $x_u$ for $I \cap S$ up to 1. 

We now break the proof into cases. 
In the first case assume that all vertices in $N(T) $ are derived some some vertex in $ B \cap S$. Then 
establish the condition in (\ref{eqn:qhinsh-proof}) as follows:
\[ \sum_{u\in S} \ell_u + |I \cap S|  = |T| \leq |N(T)| = H|B \cap S| \leq H \cdot r(S) \]
The first equality follows from the definition of $T$, the first inequality follows since $\Gamma$ matches
all vertices in $ R \cup I$, the second equality follows from the definition of $G(H)$ and the assumption of this case,
and the final inequality arrives because $B \cap S$ is an independent set in $S$.

 Otherwise, for the second case assume   $N(T)$ contains a vertex that is not derived from some  $b \in B \setminus S$.
 Thus by the definition of $G(H)$ it must be the case that there exists a $u \in S \setminus B$ such that $B - b + u \in \mathcal{I}$. Thus $B \cap S + u \in \mathcal{I}$, so 
 \begin{equation}
  |B \cap S| < r(S)
  \label{eqn:notfullrank}
 \end{equation}
 We now want to use the fact that $B \cap S$ is not full rank in $S$ to upper bound the aggregate $x$ rates for $S$ as follows:
\begin{equation}
    \sum_{u \in S} x_u = \sum_{i=1}^n \lambda_i |B_i \cap S| \leq \lambda_1(r(S)-1) + \sum_{i=2}^n \lambda_i r(S) = r(S)-\lambda_1 \leq r(S) - 1/n  \label{eqn:lambdarank}
\end{equation}
The first inequality follows from Definition \ref{defn:lambda}, the first inequality follows from equation (\ref{eqn:notfullrank}), the second equality follows by algebra, and second inequality follows from the 
assumption  that $\lambda_1$ is the largest coefficient $\lambda_i$. 
We now  establish the condition in (\ref{eqn:qhinsh-proof}) as follows:
\begin{equation*}
    \sum_{u\in S} \ell_u + |I \cap S|  \leq H \cdot \sum_{u \in S} x_u  + n \leq H(r(S) - 1/n) + n  \leq H \cdot r(S)
\end{equation*} 
The first inequality follows from the fact that by definition $\ell_u = \lfloor H \cdot x_u \rfloor$ and $I \cap S \subseteq U$,
the second inequality follows the inequality in line (\ref{eqn:lambdarank}), and the last inequality follows
from the assumption that $H \geq n^2$.
 \end{proof}

As an aside, it is relatively easy to show that the converse of Lemma \ref{lemma:qh_in_sh} holds for all $H$,
that is $M_0(H)$ is a reduction of $M_1(H)$ and thus $M_0(H)=M_1(H)$, by a straightforward application of
Hall's condition to $G(H)$.  We show this later in Claim \ref{claim:sh_in_qh}. From here on we fix $H = n^2$ and drop it from the
matroid notation. So for example $M_0$ will henceforth mean
$M_0(n^2)$.

\section{The Reduction to a Series-Parallel Matroid and the \texorpdfstring{$AR$}{AR} Process}
\label{sect:seriesparallel} 
We first explain how the algorithm constructs the series-parallel forest $T$,~\footnote{Our explanation will be self-contained, and it  is not necessary for the reader to have background knowledge on series-parallel graphs/matroids. But a curious reader can find more information at \cite{enwiki:1358933008}.  } and then define the randomized process $AR$. Our starting point for the construction of $T$ is a suitable acyclic fractional matching $F$ in $G$.
\begin{lemma}\label{lemma:acyclic_matching}
$G$ has an acyclic fractional matching $F$ that covers each element of $R$ to extent 1, each element of $Z$ to extent 1, and each $u \in U$ to extent $y_u $.
\end{lemma}

\begin{proof}
First, we show $G$ has a fractional matching $F_0$ meeting the stated covering constraints. Next, we cancel cycles in $F_0$ while maintaining the covering constraints, producing acyclic fractional matching $F$.

Let $P(M_0) = \text{conv}\{\mathbf{1}_I : I \in \mathcal{I}_0\}$. Note that $P(M_0) = \{v \in [0, 1]^U : \ell+v \in H \cdot P(M)\}$, where the subset direction is clear, and the superset direction follows by the integrality of the polytope on the right hand side. Similarly, let $P(M_1) = \text{conv}\{\mathbf{1}_I : I \in \mathcal{I}_1\}$. Note that $P(M_1) = \{v \in [0, 1]^U : v \cup R \text{ is matchable in } G\}$, where we mean that there exists a fractional matching in $G$ that matches every vertex in $R$ to extent $1$, every vertex in $Z$ to extent $1$, and every vertex $u \in U$ to extent $v_u$. Again, the equality follows because the subset direction is clear, and the superset direction follows by the integrality of the polytope on the right hand side.

\begin{claim}\label{claim:sh_in_qh}
Every independent set of $M_0$ is an independent set of $M_1$.
\end{claim}

\begin{proof}
Let $I \in \mathcal{I}_0$ be arbitrary. Then $\ell+\mathbf{1}_I \in H \cdot P(M)$. We want to show $I \in \mathcal{I}_1$, i.e. there exists a matching in $G(H)$ that matches every element in $I \cup R$. At a high level, the matroid rank constraints will supply Hall's condition, which will imply that the matching exists.

Let $T \subseteq I \cup R$ be arbitrary and $N(T) \subseteq Z$ be the neighborhood of $T$ in $G$. Let $S \subseteq U$ be the subset of elements $u \in U$ where $T$ contains a left-side vertex derived from $u$, and let $N(S) \subseteq B$ be the subset of elements $b \in B$ where $N(T)$ contains a right-hand side vertex derived from $b$. $N(S) = \{b \in B : \exists u \in S \text{ s.t. } B-b+u \in \mathcal{I}\}$. We know $|N(S)| \geq r(S)$, because for each $e \in S$ we have $e \in \text{cl}(N(S))$, so $r(S) \leq r(\text{cl}(N(S))) = r(N(S)) = |N(S)|$ \cite[Corollary 39.12a]{schrijver_book}. Thus

\[ |T| \leq \sum_{u \in S} \ell_u + |I \cap S| = \sum_{u \in S}\left(\ell_u+\mathbf{1}[u \in I \cap S]\right) \leq Hr(S) \leq H|N(S)| = |N(T)| \]

Thus $|N(T)| \geq |T|$ for all $T \subseteq I \cup R$, so by Hall's theorem, there exists a matching in $G(H)$ that matches every element in $I \cup R$. Thus $I \in \mathcal{I}_1$.
\end{proof}

Via Claim \ref{claim:sh_in_qh}, every independent set of $M_0$ is an independent set of $M_1$. Thus $P(M_0) \subseteq P(M_1)$. Because $\ell+y = Hx \in H \cdot P(M)$, we have $y \in P(M_0)$. Thus $y \in P(M_1)$. Thus $y \cup R$ is matchable in $G$, so there exists a fractional matching $F_0$ in $G$ that matches every vertex in $R$ to extent $1$, every vertex in $Z$ to extent $1$, and every vertex $u \in U$ to extent $y_u$.

Next, while $F_0$ contains a cycle $C$ in its positive support, partition the edges of $C$ into two matchings by alternating edges, and continuously increase the values on one matching while decreasing the values on the other matching at the same rate, until some edge in $C$ has value $0$. Note that the amount that $F_0$ covers each vertex does not change during this procedure. Thus upon termination, we will have an acyclic fractional matching $F$ meeting the stated covering constraints.
 \end{proof}

The algorithm then constructs a series-parallel forest $T$ from $F$ in the following manner:

\paragraph{The construction of the series-parallel forest $T$ from the fractional matching $F$:} 
First all isolated edges are removed from $F$. 
Then for each $u \in U$ that is not a leaf in $F$, give this vertex a new label $u'$, and create a new node $u$ with parent $u'$. Let $U'$ be the set of all such $u'$ created in this step. 
Then root  each tree  in the resulting forest at an arbitrary vertex in $Z$.
The nodes in $Z$ are parallel nodes and the nodes in $U' \cup R$ are series nodes. 
We then convert $T$ into a full binary tree in the standard way, by repeatedly applying an arbitrary one of the following three transformations:
\begin{itemize}
    \item If there is a node $v$ with parent $p$ and a single child $w$, then remove $v$ and make $w$ a child of $p$.
    \item If there is a parallel node $v$ with children $w_1, \ldots, w_k$ for $k \ge 3$ then create a new parallel node $v'$,
    make $w_1$ and $v'$ children of $v$, and make $w_2, \ldots, w_k$ children of $v'$. 
    \item If there is a series node $v$ with children $w_1, \ldots, w_k$ for $k \ge 3$ then create a new series node $v'$,
    make $w_1$ and $v'$ children of $v$, make and $w_2, \ldots, w_k$ children of $v'$. 
\end{itemize}

\begin{lemma}
\label{lemma:TUleaves}
The leaves of  $T$ are exactly the ground set $U$ of the matroid $M$.
\end{lemma}

\begin{proof}
We first show that the leaves of $T$ are exactly $U$ after isolated edges are removed and $U'$ is constructed, and then we show that each of the three transformations maintains the leaves of $T$.

First, observe that the leaves of $F$, which are in $R$ or $Z$, are in isolated edges of $F$. If $r \in R$ is a leaf in $F$, then its sole incident edge $e = \{r, z\}$ matches $r$ to extent $1$, and thus matches $z$ to extent $1$, so $z$ is also a leaf in $F$. Conversely, if $z \in Z$ is a leaf in $F$, then its sole incident edge $e = \{r, z\}$ matches $z$ to extent $1$, and thus matches $r$ to extent $1$, so $r$ is also a leaf in $F$. Consequently, there are no leaves in $R$ or $Z$ after isolated edges are removed. It is also clear by construction that there are no leaves in $U'$.

We now show that after isolated edges are removed and $U'$ is constructed, all of the elements $U$ are leaves. If $u \in U$ is a leaf in $F$, then we know that the other endpoint $z \in Z$ on the unique edge $e = \{u, z\}$ cannot be a leaf in $F$, so $e$ is not an isolated edge. Thus $u$ remains a leaf in $T$. Further, if $u \in U$ is not a leaf in $F$, then it is automatically made a leaf because of its replacement with $u'$.

Lastly, it is easy to check that each of the three transformations changes the children of nodes in $T$, but not \emph{whether} a node has children (i.e. whether a node is a leaf). Thus the leaves of the full binary forest $T$ are exactly $U$.
 \end{proof}

  We now inductively extend the probabilities $y_u$ on the leaves of $T$ to the internal nodes.

\begin{definition}
For each internal node $v \in T$,  with children $w_1$ and $w_2$,  we define $y_v$  to be
$y_{w_1} + y_{w_2}$ if $v$ is a parallel node, and to be $y_{w_1} + y_{w_2}  - 1$ 
if $v$ is a series node. 
\end{definition}

\begin{lemma} 
\label{lemma:yvin01}
    For all nodes $v \in T$ it is the case that $y_v \in [0, 1]$.
\end{lemma}

\begin{proof}
Let $T_0$ be the forest computed before repeatedly performing the three transformations. First, we show that the claim holds on the forest $T_0$, under a general rule for constructing the $y_v$ values, and then we show that each of the three transformations preserve the $y_v$ values.

For an edge $e = \{v, p\}$ between vertex $v$ and its parent $p$ in $T_0$, let $F_{vp}$ be the value on edge $e$ in the acyclic fractional matching $F$, where we set $F_{u'p} = F_{up}$ for a vertex $u' \in U'$ and its corresponding vertex $u \in U$. For a vertex $v \in T_0$, let $y_v' = y_v$ if $v$ is a leaf in $T_0$. Otherwise, $v$ has children $w_1, w_2, \dots, w_k \in T_0$, and recursively define $y_v' = \sum_{i=1}^k y'_{w_i}$ if $v$ is a parallel node (equivalently $v \in Z$), and $y_v' = \sum_{i=1}^k y'_{w_i} - (k-1)$ if $v$ is a series node (equivalently $v \in R \cup U'$). We show by induction from the leaves upward that
\[
y_v'=
\begin{cases}
1-F_{vp} &\text{if $v \in Z$ and $v$ is a nonroot vertex}\\
1 &\text{if $v \in Z$ and $v$ is a root vertex} \\
F_{vp} &\text{if $v \in R \cup U'$}
\end{cases}
\]

First, if $v \in Z$ and $v$ is a nonroot vertex, then

\[ y_v = \sum_{i=1}^k y_{w_i} = \sum_{w_i \in R \cup U'} F_{w_i v} + \sum_{w_i \in U} y_{w_i} = \sum_{w_i \in R \cup U'} F_{w_i v} +  \sum_{w_i \in U} F_{w_iv} = 1-F_{vp} \]

The first equality follows by the parallel recurrence. The second equality follows by the inductive hypothesis $y_{w_i} = F_{w_i v}$ for all $w_i \in R \cup U'$. The third equality follows by $y_{w_i} = F_{w_i v}$ for all $w_i \in U$, because $e = \{w_i, v\}$ is the sole edge in $F$ which is matching $w_i$, so $e$ must match $w_i$ to extent $y_{w_i}$. The fourth equality follows because $v \in Z$ is matched to extent $1$. 

Second, if $v \in Z$ and $v$ is a root vertex, then the same reasoning gives $y_v = 1$.

Third, if $v \in R$, then

\[ y_v = \sum_{i=1}^k y_{w_i} - (k-1) = \sum_{i=1}^k \left(1-F_{w_i v}\right) - (k-1) = 1-\sum_{i=1}^k F_{w_i v} = F_{vp} \]

The first equality follows by the series recurrence. The second equality follows by the inductive hypothesis $y_{w_i} = 1 - F_{w_i v}$ for all $w_i \in Z$. The third equality follows by simplification. The fourth equality follows because $v \in R$ is matched to extent $1$. Lastly, if $v \in U'$, then let $w_k$ be the unique child of $v$ in $U$, giving

\[ y_v = \sum_{i=1}^k y_{w_i} - (k-1) = \sum_{i=1}^{k-1} \left(1-F_{w_i v}\right) + y_{w_k} - (k-1) = y_{w_k} - \sum_{i=1}^{k-1} F_{w_i v} = F_{vp} \]

The first equality follows by the series recurrence. The second equality follows by the inductive hypothesis $y_{w_i} = 1 - F_{w_i v}$ for all $w_i \in Z$. The third equality follows by simplification. The fourth equality follows because $u \in U$ is matched to extent $1$. 

Thus because $F_{pv} \in [0, 1]$ for each edge $e = \{v, p\}$ between vertex $v$ and its parent $p$ in $T_0$, and because $y_u' \in [0, 1]$ for all leaves $u \in U$, we have $y_v' \in [0, 1]$ for all $v \in T_0$.

Next, we show that when $T_0$ is altered under the three transformations, the $y_v'$ values are still preserved under the general series and parallel recurrence rules, and new vertices $v'$ which are created satisfy $y_{v'} \in [0, 1]$. Because the general rules are the same as the binary tree rules for $k=2$, this will complete the proof.

First, if a node $v$ has a single child $w$, then $y_v = y_w$ regardless of whether $v$ is a parallel node or a series node. Thus replacing $v$ with $w$ does not change the value $y_p$ of the parent $p$ of $v$. Second, if a node $v$ is a parallel node with children $w_1, w_2, \dots, w_k$ for $k \geq 3$, then we replace $v$'s children with $w_1, v'$ where $v'$ is parallel and has children $w_2, w_3, \dots, w_k$. Thus we have 

\[ y_v = y_{w_1} + y_{v'} = y_{w_1} + \sum_{i=2}^k y_{w_i} = \sum_{i=1}^k y_{w_i} \] 

as before, and further $0 \leq y_{v'} \leq y_v \leq 1$ so $y_{v'} \in [0, 1]$. Lastly, if a node $v$ is a series node with children $w_1, w_2, \dots, w_k$ for $k \geq 3$, then we replace $v$'s children with $w_1, v'$ where $v'$ is series and has children $w_2, w_3, \dots, w_k$. Thus we have

\[ y_v = y_{w_1} + y_{v'} - 1 = y_{w_1} + \sum_{i=2}^k y_{w_i} - (k-2) - 1 = \sum_{i=1}^k y_{w_i} - (k-1)\]

as before, and further $0 \leq y_v \leq y_{v'} \leq 1$ so $y_{v'} \in [0, 1]$. Thus $y_v \in [0, 1]$ for all $v \in T$, completing the proof.
\end{proof}

\begin{definition}
For each internal node $v \in T$, with children $w_1$ and $w_2$, we define $\alpha_{w_i}$ for each child of $v$ as follows:
    \begin{itemize}
        \item If $v$ is a parallel node and $y_v \ne 0$ then $\alpha_{w_i} = \frac{y_{w_i}}{y_v}$.
        \item If $v$ is a parallel node and $y_v = 0$ then $\alpha_{w_i} = \frac{1}{2}$.
        \item If $v$ is a series node and $1-y_v \ne 0$ then $\alpha_{w_i} = \frac{1-y_{w_i}}{1-y_v}$.
        \item If $v$ is a series node and $1-y_v = 0$ then $\alpha_{w_i} = \frac{1}{2}$.
    \end{itemize}
\end{definition}

We define a randomized process $AR$ on each tree in the forest $T$ that we will show generates a basis $I$ of matroid $M_1$, 
and that matches the marginal
probabilities $y_u$ for each leaf $u \in U$. 

\paragraph{The $AR$ Process:}
In this process  each internal node $v \in T$ receives a bit from its parent (other than the roots which always receive a 1 bit) and passes along bits to its children in the following manner:
\begin{itemize}
    \item If $v$ is a parallel node and $v$ receives a 0 bit from its parent then $v$ passes a 0 bit to each child. 
    \item If $v$ is a parallel node and $v$ receives a 1 bit from its parent then $v$ passes a 1 bit to each child $w$ of $v$ with probability $\alpha_w$, and passes a 0 bit to the other child. 
     \item If $v$ is a series node and $v$ receives a 1 bit from its parent then $v$ passes a 1 bit to each child. 
    \item If $v$ is a series node and $v$ receives a 0 bit from its parent then $v$ passes a 0 bit to each child $w$ of $v$ with probability $\alpha_w$, and passes a 1 bit to the other child. 
\end{itemize}

\begin{lemma} \label{lemma:ARprobyv}
The $AR$ process is well-defined, and each vertex $v \in T$ receives bit $1$ with probability $y_v$.
\end{lemma}

\begin{proof}
First, the transition probabilities are well-defined. At a parallel node $v$ with children $w_1$ and $w_2$ s.t. $y_v>0$, we have

\[ \alpha_{w_1}+\alpha_{w_2} = \frac{y_{w
_1}}{y_v} + \frac{y_{w_2}}{y_v} = \frac{y_{w_1} + y_{w_2}}{y_v} = 1 \]

and at a series node $v$ with children $w_1$ and $w_2$ s.t. $y_v < 1$, we have

\[ \alpha_{w_1}+\alpha_{w_2} = \frac{1-y_{w
_1}}{1-y_v} + \frac{1-y_{w_2}}{1-y_v} = \frac{1-(y_{w_1} + y_{w_2}-1)}{1-y_v} = 1 \]

We now induct down the tree. Each root $r$ receives bit $1$ with probability $1=y_r$, as shown in the proof of Lemma \ref{lemma:yvin01}. Suppose a vertex $v$ receives bit $1$ with probability $y_v$, and let $w$ be a child of $v$.

If $v$ is a parallel node and $y_v>0$, then
\begin{equation*}
\Pr[w\text{ receives }1]
=y_v\alpha_w
=y_v\frac{y_w}{y_v}
=y_w.
\end{equation*}

If $v$ is a parallel node and $y_v = 0$, then $y_w = 0$. Further, by induction $v$ always receives bit $0$, so $w$ always receives bit $0$. Thus $\Pr[w\text{ receives 1}] = y_w$.

If $v$ is a series node and $y_v<1$, then
\begin{equation*}
\Pr[w\text{ receives }0]
=(1-y_v)\alpha_w
=(1-y_v)\frac{1-y_w}{1-y_v}
=1-y_w.
\end{equation*}

so $\Pr[w\text{ receives }1] = y_w$. If $v$ is a series node and $y_v = 1$, then $y_w = 1$. Further, by induction $v$ always receives bit $1$, so $w$ always receives bit $1$. Thus $\Pr[w\text{ receives 1}] = y_w$.

Thus the $AR$ process is well-defined, and every vertex $v \in T$ receives bit $1$ with probability $y_v$.
\end{proof}

\begin{definition}
In the set system $M_2 = (U, \mathcal{I}_2)$ a set $I \subseteq U$ is in $\mathcal{I}_2$ if and only if 
the probability that $I$ are exactly the leaves that receive 1 bits in the $AR$ process is positive. 
\end{definition}

\begin{lemma} \label{lemma:M2M1}
Every independent set of $M_2$ is an independent set of $M_1$.
\end{lemma}

\begin{proof}
Let $I \in \mathcal{I}_2$ be arbitrary, and fix some realization of the $AR$ process where $I \subseteq U$ is exactly the set of leaves which receive bit $1$. Let $Z_T = Z \cap T$ and $R_T = R \cap T$. For each $z \in Z_T$, if $z$ received bit $0$ then match $z$ to its parent, and else $z$ received bit $1$, so match $z$ to its unique child which also received $1$ (a unique child exists because $z$ is a parallel node). Then modify the matching by taking each $z \in Z_T$ matched to some $u' \in U$ and instead matching $z$ to the corresponding $u \in U$. Lastly, for each $z \in Z \setminus Z_T$, match $z$ to its unique neighbor $r \in R \setminus R_T$ in $F$ using the isolated edge $\{r, z\}$. Let $N$ be the resulting matching. We now show that $N$ is a matching in $G$ that matches every element in $I \cup R$, proving that $I \in \mathcal{I}_1$.

Let $r \in R$ be arbitrary. If $r \in R \setminus R_T$, then $r$ is matched to its unique neighbor $z \in Z \setminus Z_T$ in $F$ using the isolated edge $\{r, z\}$. Otherwise $r \in R_T$. If $r$ received bit $0$, then it passes bit $0$ to some child $z \in Z_T$, and then $r, z$ are matched. Otherwise $r$ received bit $1$, so it received bit $1$ from its parent $z \in Z_T$, and then $r, z$ are matched. Thus $R$ is matched.

Let $u \in I$ be arbitrary. Then $u$ received bit $1$. If the parent of $u$ is some $z \in Z_T$, then $z$ received bit $1$, and $u, z$ are matched. Otherwise the parent of $u$ is the corresponding $u' \in U'$. If $u'$ received bit $0$, then $u'$ passes $0$ to exactly one child $z \in Z_T$, and $u', z$ are originally matched, so $u, z$ are matched in the end. Otherwise $u'$ received bit $1$, so the parent $z \in Z_T$ received bit $1$, and $u', z$ are originally matched, so $u, z$ are matched in the end. Thus $I$ is matched.

Let $u \in U \setminus I$ be arbitrary. Then $u$ received bit $0$. If the parent of $u$ is some $z \in Z_T$, then if $z$ received bit $0$ then $z$ is matched to its parent, and if $z$ received bit $1$ then it is matched to its unique child which also received bit $1$. Thus $u$ is not matched. Otherwise the parent of $u$ is the corresponding $u' \in U'$, which also received bit $0$. But then $u'$ passed bit $1$ to all of its other children $z \in Z_T$, which will each match to their children who also received bit $1$, and for the same reasoning as before, $u'$ is not matched to its parent because it received bit $0$. Thus $u'$ is not originally matched, so $u$ is not matched.

Thus $N$ matches exactly $I \cup R$ in $G$, so $I \in \mathcal{I}_1$.
 \end{proof}

\section{Chunking the Series-Parallel Forest \texorpdfstring{$T$}{T}}
\label{sect:chunkingT}

In this section, we explain how our algorithm recursively constructs the chunks
from $T$ that will form the chunk tree.

\begin{definition}
For each internal node $v \in T$,  with children $w_1$ and $w_2$, 
we define the favorite child of $v$ to be $w_1 $ if $\alpha_{w_1} \ge \alpha_{w_2}$,
and to be $w_2$ otherwise.
 \end{definition}

 \paragraph{Chunk Forming Algorithm:}
 Let $\rho $ be a parameter that we will set later to be $2/3$. 
 Let $v_1$ be a root of $T$. If the favorite child $w$ of $v_1$ is a leaf, or has $\alpha_w < \rho$, then let the chunk $C = v_1$. Otherwise, let $v_1, ...., v_{k+1}$ be the ancestor to descendant path in $T$ 
 where each $v_i$ is the favorite of its parent $v_{i-1}$, and 
 \begin{itemize}
     \item $\prod_{i=2}^{k} \alpha_{v_i} \ge \rho$, and
     \item $v_{k+1}$ is a leaf or $\prod_{i=2}^{k+1} \alpha_{v_i} < \rho$.
 \end{itemize}
Then the path $C=v_1, ...., v_{k}$ becomes a  chunk, 
 and the process continues recursively on each tree in the forest $T-C$ (other than trees that consist of a single leaf).

\begin{definition}\label{defn:condprob}
    Let $C = v_1, \ldots, v_k$ be a chunk of the forest $T$, and $L(C) = w_1, \ldots, w_{k+1}$ be the children of the nodes in $C$
    where $w_i$ is the child of $v_i$ not in $C$ for all $i \in [k-1]$,  $w_{k+1}$ is the favorite child of $v_k$, and
    $w_k$ is the other child of $v_k$.
   For each $w_i$:
   \begin{itemize}
       \item Define $y^{0}_{w_i}$ to be the probability that node $w_i$ receives a 1 bit in the $AR$ process on $T$
   conditioned on the fact that $v_1$ received a 0 bit in the process.
   \item Define $y^{1}_{w_i}$ to be the probability that node $w_i$ receives a 1 bit in the $AR$ process on $T$
   conditioned on the fact that $v_1$ received a 1 bit in the process. 
   \end{itemize} 
\end{definition}

\begin{lemma} \label{lemma:computey} 
Consider the setting from Definition \ref{defn:condprob}.  Let $w = w_i$ for some $i \in [k+1]$. Then $y_w^0$ and $ y_w^1$ is the unique solution to the two equations
\begin{align}
y_w &= \left(1-y_{v_1}\right)y_w^0 + y_{v_1}y_w^1 \\
\alpha(i) &= y_w^1-y_w^0
\end{align}
where $\alpha(i)= \alpha_{w_1}$ if $i=1$,  $\alpha(i)= \prod_{j=2}^i \alpha_{v_j} \cdot \alpha_{w_i}$ if $i \in [2, k]$, and 
$\alpha(i)=  \prod_{j=2}^k \alpha_{v_j} \cdot \alpha_{w_{k+1}}$ if $i=k+1$.
\end{lemma}

\begin{proof}
First, we show that $y_w = (1-y_v)y_w^0+y_vy_w^1$. This is because $y_w$ is the probability that $w$ receives bit $1$ and $y_v$ is the probability that $v$ receives bit $1$ in the $AR$ process (Lemma \ref{lemma:ARprobyv}), so we can compute $y_w$ by conditioning on whether $v$ receives bit $0$ or bit $1$, and multiplying by the probabilities of these events $1-y_v$ and $y_v$, respectively.

Next, we show that $\alpha(i) = y_w^1-y_w^0$. Define $y_{v_j}^{\gamma}$ as the probability that $v_j$ receives bit $1$ given that $v_1 = v$ receives bit $\gamma \in \{0, 1\}$ in the $AR$ process for $j \in [k]$. Define

\[
\alpha'(h)=
\begin{cases}
1 &\text{if $h=1$}\\
\prod_{j=2}^h \alpha_{v_h} &\text{if $h \in [2, k]$}
\end{cases}
\]

We now show by induction that $y_{v_j}^1 - y_{v_j}^0 = \alpha'(j)$ and $y_{w_j}^1 - y_{w_j}^0 = \alpha(j)$ for all $j \in [k]$, and that $y_{w_{k+1}}^1 - y_{w_{k+1}}^0 = \alpha(k+1)$. First, note $y_{v_1}^1 = 1$, $y_{v_1}^0 = 0$, and $\alpha'(1) = 1$, so we have $y_{v_1}^1 - y_{v_1}^0 = \alpha'(1)$. Next, assume $y_{v_j}^1 - y_{v_j}^0 = 
\alpha'(j)$ for some $j \in [k-1]$. If $v_j$ is a parallel node, then $w_j$ receives bit $1$ if and only if $v_j$ receives bit $1$ and passes bit $1$ to $w_j$. Thus $y_{w_j}^1 = y_{v_j}^1 \cdot \alpha_{w_j}$ and $y_{w_j}^0 = y_{v_j}^0 \cdot \alpha_{w_j}$, so $y_{w_j}^1-y_{w_j}^0 = \left(y_{v_j}^1-y_{v_j}^0\right)\alpha_{w_j} = \alpha'(j)\alpha_{w_j} = \alpha(j)$. Similarly, if $v_j$ is a series node, then $w_j$ receives bit $0$ if and only if $v_j$ receives bit $0$ and passes bit $0$ to $w_j$. Thus $1-y_{w_j}^1 = \left(1-y_{v_j}^1\right)\alpha_{w_j}$ and $1-y_{w_j}^0 = \left(1-y_{v_j}^0\right)\alpha_{w_j}$, so $y_{w_j}^1-y_{w_j}^0 = \left(1-y_{w_j}^0\right) - \left(1-y_{w_j}^1\right) = \left(\left(1-y_{v_j}^0\right)-\left(1-y_{v_j}^1\right)\right)\alpha_{w_j} = \left(y_{v_j}^1 - y_{v_j}^0\right)\alpha_{w_j} = \alpha'(j)\alpha_{w_j} = \alpha(j)$. The same reasoning for $v_{j+1}$ gives $y_{v_{j+1}}^1-y_{v_{j+1}}^0 = \alpha'(j)\alpha_{v_{j+1}} = \alpha'(j+1)$. This completes the induction, and it remains to show the equation holds for $w_k$ and $w_{k+1}$. The same reasoning gives $y_{w_k}^1-y_{w_k}^0 = \alpha'(k)\alpha_{w_k} = \alpha(k)$ and $y_{w_{k+1}}^1-y_{w_{k+1}}^0 = \alpha'(k)\alpha_{w_{k+1}} = \alpha(k+1)$, completing the proof.
\end{proof}

 We are now ready to describe our algorithm $A$ for producing the sequence of fills.
 This algorithm uses two subroutines for each chunk $C$, a 0 parity chunk algorithm, and a 1 parity chunk algorithm.
 Each subroutine when called will output a bit for each child of a node in $C$.

 \paragraph{The Fill Algorithm $A$:}
 Initially, for each root of $T$, the 1 parity chunk algorithm is called for the chunk $C$ containing the root, which outputs a bit for each child $w$ of a node in $C$. 
 The algorithm then iteratively considers chunks $C$ whose root $v$ has received a bit, but  no parity chunk algorithm
 has been called yet for $C$. If $v$ received a 0 bit then the 0 parity chunk algorithm for $C$ is called,
 else ($v$ received a 1 bit)  the 1 parity chunk algorithm for $C$ is called. 
In each case the parity chunk algorithm outputs a bit for each child $w$ of a node in $C$. 
 This process terminates when a parity chunk algorithm has been called for each chunk. The next
 fill is then exactly the leaves $I$ that received a 1 bit.

\section{The Chunk Algorithm}
\label{sect:chunkalgorithm}

The goal of this section is to describe the parity chunk algorithm for a particular parity $\gamma$ and a chunk $ C = v_1, \ldots v_k$.
 Each time each algorithm is called, it will compute a subset of the collection $L(C)=w_1, \ldots, w_{k+1}$ of children of $C$, send $1$ bits to the children in this subset, and send $0$ bits to the children not in this subset. Here the parent of $w_i$, $i \in [1, k]$, is $v_i$, and $w_{k+1}$ is the favorite child of $v_k$. 
The $\gamma$ parity chunk algorithm maintains a default base $B^\gamma \subseteq L(C)$,
and each computed subset will be of the form $B^\gamma - w_i + w_j$. 

The $\gamma$ parity chunk algorithm maintains a unit clock for the chunk $C$, which we call the chunk clock, and a unit clock for each vertex $w_i \in L(C)$, which we call the vertex clocks. Each clock will only advance when called by the $\gamma$ parity chunk algorithm. First, the unit clock for the chunk $C$ is called, which will return a selected element of $L(C)$.
If the selected element is in $B^\gamma$, then it will become $w_i$, and otherwise it will become $w_j$. 
If the selected element $w_i \in B^\gamma$ then the unit clock for $w_i$ will be called to select $w_j$. 
If the selected element $w_j \notin B^\gamma$ then the unit clock for $w_j$ will be called to select $w_i$. 
So it remains to describe how the $\gamma$ parity chunk algorithms initially compute the default bases, and set up the unit clocks.

\subsection{The Default Bases}

The default bases are conceptually formed by pushing the parity bit $\gamma$ down the chunk, ending
with the favorite child $w_{k+1}$ of $v_k$. This results in the following default bases:
\begin{itemize}
    \item 
Let $B^1 = b_1^1, \ldots, b_{\beta(1)}^1$ be the elements of $L(C)$,  ordered by increasing depth, whose parents in $C$ are series nodes, followed by $w_{k+1}$ if $v_k$ is a parallel node. 
\item
Let $E^1 = e_1^1, \ldots, e_{\eta(1)}^1$ denote the elements of $L(C)$, ordered by increasing depth, whose parents in $C$ are parallel nodes, excluding $w_{k+1}$ if $v_k$ is a parallel node. 
\item 
Let $B^0 = b_1^0, \ldots, b_{\beta(0)}^0$ be the elements of $L(C)$, ordered by increasing depth, whose parents in $C$ are series nodes, excluding $w_{k+1}$ if $v_k$ is a series node.
\item
Let $E^0 = e^0_1, \ldots, e^0_{\eta(0)}$ denote the elements of $L(C)$, ordered by increasing depth, whose parents in $C$ are parallel nodes, followed by $w_{k+1}$ if $v_k$ is a series node. 
\end{itemize}

Before setting up the unit clocks, we need to prove some facts about the relationship between various probabilities
that nodes receive a 1 bit during the $AR$ process. For ease of notation, we write $y_{e_i}^{\gamma}$ to mean $y_{e_i^{\gamma}}^{\gamma}$ and $y_{b_i}^{\gamma}$ to mean $y_{b_i^{\gamma}}^{\gamma}$.

\begin{lemma} \label{lemma:betaeta}
For $\gamma \in \{0, 1\}$ it is the case that $\sum_{i=1}^{\eta(\gamma)} y_{e_i}^{\gamma} = \sum_{i=1}^{\beta(\gamma)} (1-y_{b_i}^\gamma)$.
\end{lemma}

\begin{proof}
    This follows because the number of elements in $L(C)$ receiving a 1 bit in
    the $AR$ process is exactly $\gamma$ plus the number of series nodes in $C$, regardless of the random events internal to $C$. This quantity is exactly $\beta(\gamma)$. Thus $\sum_{i=1}^{\eta(\gamma)} y_{e_i}^{\gamma} + \sum_{i=1}^{\beta(\gamma)} y_{b_i}^{\gamma} = \beta(\gamma)$, so $\sum_{i=1}^{\eta(\gamma)} y_{e_i}^{\gamma} = \sum_{i=1}^{\beta(\gamma)} (1-y_{b_i}^{\gamma})$.
\end{proof}

\begin{lemma} \label{lemma:nearbasis}
    For $\gamma \in \{0, 1\}$ it is the case that $\sum_{i=1}^{\eta(\gamma)}  y^\gamma_{e_i } \le 1$.
\end{lemma}

\begin{proof}
First, consider $\gamma = 1$. An element $e_i \in E^1$ receives a bit $1$ if and only if its parent $v_j \in C$ receives bit $1$, and $v_j$ passes bit $1$ to $e_i$, because $v_j$ is a parallel node. Thus $y_{e_i}^1 \leq \alpha_{e_i} = 1-\alpha_{v_{j+1}}$ if $j \in [k-1]$, and $1-\alpha_{w_{k+1}}$ if $j = k$. Thus

\[ \sum_{i=1}^{\eta(1)} y_{e_i}^1 \leq \sum_{j=2}^k \left(1-\alpha_{v_j}\right) + \left(1-\alpha_{w_{k+1}}\right) \leq -\log \prod_{j=2}^k \alpha_{v_j} + \frac{1}{2} \leq \frac{1}{2}-\log \rho \leq 1 \]

The first inequality follows by the stated upper bounds on $y_{e_i}^1, i \in [\eta(1)]$. The second inequality follows by $1-\alpha \leq -\log \alpha$ for all $\alpha > 0$, and because $w_{k+1}$ is the favorite child of $v_k$, so $\alpha_{w_{k+1}} \geq \frac{1}{2}$. The third inequality follows by $\prod_{j=2}^k \alpha_{v_j} \geq \rho$. The fourth inequality follows by the choice $\rho=2/3$.

Next, consider $\gamma = 0$. An element $b_i \in B^0$ receives a bit $0$ if and only if its parent $v_j \in C$ receives bit $0$, and $v_j$ passes bit $0$ to $b_i$, because $v_j$ is a series node. Thus $1-y_{b_i}^0 \leq \alpha_{b_i} = 1-\alpha_{v_{j+1}}$ if $j \in [k-1]$, and $1-\alpha_{w_{k+1}}$ if $j = k$. Thus

\[ \sum_{i=1}^{\eta(0)} y_{e_i}^0 =  \sum_{i=1}^{\beta(0)} \left(1-y_{b_i}^0\right) \leq \sum_{j=2}^k \left(1-\alpha_{v_j}\right) + \left(1-\alpha_{w_{k+1}}\right) \leq 1 \]

The first inequality follows by Lemma \ref{lemma:betaeta}. The second inequality follows by the stated upper bounds on $1-y_{b_i}^0, i \in [\beta(0)]$. The third inequality follows by the same reasoning as before. 
\end{proof}

Lemma \ref{lemma:betaeta} follows immediately from the fact that the number of 1 bits passed to a node in $L(C)$ depends only on the parity bit $\gamma$. This will allow us to conceptually perfectly match elements in the default base with elements not in the default base. Lemma \ref{lemma:nearbasis} show that the aggregate probability
that $AR$ sends a 1 to  elements in $L(C)$ that are not in
the default base is at most 1. This is what enables the algorithm to always output 1 bits to a subset of $L(C)$ that is within one swap of the appropriate parity default base.

\subsection{Setting Up the Chunk Clocks}

\begin{figure}[htbp]
\centering
\includegraphics[width=0.9\linewidth]{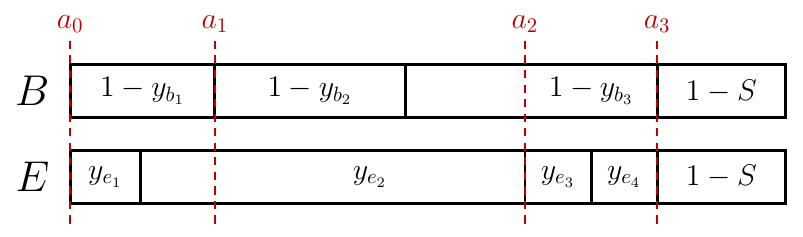}
\caption{An example instance for constructing the chunk and vertex clocks. See Figure  \ref{fig:ChunkInternalClock} for a illustration of the resulting clocks. }
\label{fig: ChunkClock}
\end{figure}

The construction of the clocks is not conceptually difficult, but the notation required for a formal
definition is somewhat painful. The reader is encouraged to refer frequently to the example shown in Figure \ref{fig: ChunkClock}. Once one understands the example in this figure, both the algorithm design and analysis should be clear. 
We first need some definitions. Let $S^\gamma = \sum_{i=1}^{\beta(\gamma)} (1- y^\gamma_{b_i})  = \sum_{i=1}^{\eta(\gamma)}  y^\gamma_{e_i } $.
Let $Q_{b,i}^\gamma = \sum_{j=1}^i ( 1 - y^\gamma_{b_j})$, and $Q_{e,i}^{\gamma} = \sum_{j=1}^i y^\gamma_{e_i}$.
To reduce notational complexity a bit, we are now going to drop the reference to $\gamma$ in the variables.

The algorithm first computes intervals $(a_0 = 0 , a_1), (a_1, a_2), ..., (a_{\ell-1}, a_{\ell} = S)$ as follows:
\begin{itemize}
    \item $a_0 = 0$. 
    \item $a_i$ is maximum (but at most $S$) so that  $(a_{i-1}, a_i)$ intersects only one interval of the form 
    $(Q_{b,j-1} , Q_{b, j} )$ or only intersects one interval of the form
  $(Q_{e, j-1} , Q_{e, j} )$. 
\end{itemize}
The $\gamma$ parity chunk clock then contains $ \ell+1$ hands, with associated probabilities:
\[a_1 , (a_2 - a_1), \ldots,(a_{\ell-1} - a_{\ell -2}), (S - a_{\ell-1}), 1-S\]

If the chunk clock outputs the $(\ell+1)$'st hand, which is not associated with an interval, then the algorithm outputs $B$. Otherwise, the chunk clock outputs the $k$'th hand corresponding to interval $(a_{k-1}, a_k)$ for some $k \in [\ell]$, and the vertex clock for $(a_{k-1}, a_k)$ is called.

\subsection{Setting Up the Vertex Clocks}

The algorithm maintains a vertex clock for each interval of the form $(a_{k-1}, a_k)$ for $k \in [\ell]$.
First  consider the case that interval $(a_{k-1}, a_{k})$ intersects only one interval of the form
  $(Q_{b, j-1} , Q_{b, j} )$. We will call $b_j$ the dominant vertex for interval $(a_{k-1}, a_k)$.
  Let $$(Q_{e,c-1}, Q_{e, c}), \ldots, (Q_{e, d-1}, Q_{e, d})$$ be the intervals of this 
  form that intersect $(a_{k-1}, a_{k})$. Then the algorithm initializes
 a vertex clock   for the interval $(a_{k-1}, a_{k})$ with possible outcomes $e_c, \ldots e_d$, which we call subordinate vertices.
The  probability of outcome $e_i$ is the fraction of the interval $(a_{k-1}, a_k)$ covered by the interval $(Q_{e, i-1}, Q_{e, i})$,
that is 
\[ \frac{ |  (Q_{e, i-1}, Q_{e, i})  \cap (a_{k-1}, a_{k}) |}{a_k - a_{k-1}}\]
If the vertex clock outputs $e_i$, then the algorithm outputs $B-b_j+e_i$.

Now consider the case that $(a_{k-1}, a_{k})$ intersects only one interval of the form
  $(Q_{e, j-1} , Q_{e, j} )$. We will call $e_j$ the dominant vertex for interval $(a_{k-1}, a_k)$.
 Let $$(Q_{b,c-1}, Q_{b, c}), \ldots, (Q_{b, d-1}, Q_{b, d})$$ be the intervals of this 
  form that intersect $(a_{k-1}, a_{k})$. Then the algorithm initializes
 a vertex clock   for the interval $(a_{k-1}, a_{k})$ with possible outcomes $b_c, \ldots b_d$, which we call the subordinate vertices.
The  probability of outcome $b_i$ is the fraction of the interval $(a_{k-1}, a_k)$ covered by the interval  $(Q_{b, i-1}, Q_{b, i})$,
that is 
\[ \frac{ |  (Q_{b, i-1}, Q_{b, i})  \cap (a_{k-1}, a_{k}) |}{a_k - a_{k-1}}\]
If the vertex clock outputs $b_i$, then the algorithm outputs $B-b_i+e_j$.

\subsection{Correctness}
 
The fact that the construction of the chunk clocks and vertex clocks are well-defined follows from
Lemma \ref{lemma:betaeta} and Lemma \ref{lemma:nearbasis}. It remains to show that 
if the chunk algorithm outputs $B^{\gamma}$ or $B^\gamma - b_i^\gamma + e_j^\gamma$, then this output can be implemented into a feasible execution of the $AR$ process.

\begin{lemma}\label{lemma:chunk_feasible}
For $\gamma \in \{0, 1\}$ it is the case that for the output $\bar{B}^{\gamma}$ of the $\gamma$ parity chunk algorithm, where $\bar{B}^{\gamma} = B^{\gamma}$ or $\bar{B}^{\gamma} = B^{\gamma}-b_i^{\gamma}+e_j^{\gamma}$, there exists a feasible execution of the $AR$ process where the nodes $\bar{B}^{\gamma}$ receive $1$ bits and the nodes $L(C) \setminus \bar{B}^{\gamma}$ receives $0$ bits. 
\end{lemma}

\begin{proof}
First, consider $\gamma = 1$. The nodes $B^1$ could receive $1$ bits and the nodes $L(C) \setminus B^1 = E^1$ could receive $0$ bits if the $\gamma = 1$ bit is passed down the entire chunk by definition of $B^1, E^1$. Now consider the nodes $B^1-b_i^1+e_j^1$ output by the chunk algorithm. We claim that $e_j^1$ appears higher in the tree than $b_i^1$, i.e. $e_j^1 = w_h$ and $b_i^1 = w_g$ for $h < g$. If the claim holds, then the nodes $B^1-b_i^1+e_j^1$ could receive $1$ bits and the nodes $L(C) \setminus (B^1-b_i^1+e_j^1) = E^1 - e_j^1 + b_i^1$ could receive $0$ bits in the following execution of the $AR$ process: the $\gamma=1$ bit is passed down the chunk, flipping at the parent of $e_j^1$ (sending a $1$ bit to $e_j^1$, and a $0$ bit to its sibling), passing the $0$ bit down the chunk, flipping back at the parent of $b_i^1$ (sending a $0$ bit to $b_i^1$, and a $1$ bit to its sibling), and passing the $1$ bit down the rest of the chunk.

\begin{claim}
If the $1$ parity chunk algorithm outputs $B^1-b_i^1+e_j^1$, then $e_j^1$ appears higher in the tree than $b_i^1$.
\end{claim}

\begin{proof}
We first show that if $Q_{b, c}^1 > Q_{e, d-1}^1$ for some $c \in [\beta(1)]$ and $d \in [\eta(1)]$, then $e_d^1$ is above $b_c^1$ in the tree. If $b_c^1 = w_{k+1}$, then the claim trivially holds. Otherwise, note that in every execution of the $AR$ process on the chunk with input bit $\gamma = 1$, the nodes $e^1 \in E^1$ which receive bit $1$ and the nodes $b^1 \in B^1 \setminus \{w_{k+1}\}$ which receive bit $0$ alternate going down the chunk, starting with a node $e^1 \in E^1$. Thus for any prefix of the chunk $C_h = v_1, v_2, \dots, v_h$ with children $L(C_h) = w_1, w_2, \dots, w_h$, the expected number of children in $L(C_h) \cap E^1$ receiving bit $1$ is at least the expected number of children in $L(C_h) \cap B^1$ receiving bit $0$. Formally, let $e_d$ be the last child in $L(C_h) \cap E^1$ and $b_c$ be the last child in $L(C_h) \cap B^1$; then we have $Q_{b, c}^1 \leq Q_{e, d}^1$. Now consider the prefix of the chunk $C_h = v_1, v_2, \dots, v_h$ where $b_c^1= v_h$. If $e_d^1$ is below $b_c^1$ in the tree, then $Q_{b, c}^1 \leq Q_{e, d-1}^1$. Thus if $Q_{b, c}^1 > Q_{e, d-1}^1$, $e_d^1$ is above $b_c^1$ in the tree.

Next, we establish the claim. It suffices to show that $Q_{b, i}^1 > Q_{e, j-1}^1$. If $e_j^1$ is the dominant vertex in its interval $(a_{k-1}, a_k)$, then the interval starts at $a_{k-1} \geq Q_{e, j-1}^1$, so we have $Q_{b, i}^1 > a_{k-1} \geq Q_{e, j-1}^1$. If $b_i^1$ is the dominant vertex in its interval $(a_{k-1}, a_k)$, then the interval ends at $a_k = Q_{b, i}^1$, so we have $Q_{b, i}^1 \geq Q_{e, j}^1 > Q_{e, j-1}^1$.
\end{proof}

Next, consider $\gamma = 0$. The nodes $B^0$ could receive $1$ bits and the nodes $L(C) \setminus B^0 = E^0$ could receive $0$ bits if the $\gamma = 0$ bit is passed down the entire chunk by definition of $B^0, E^0$. Now consider the nodes $B^0-b_i^0+e_j^0$ output by the chunk algorithm. We claim that $b_i^0$ appears higher in the tree than $e_j^0$, i.e. $b_i^0 = w_h$ and $e_j^0 = w_g$ for $h < g$. If the claim holds, then the nodes $B^0-b_i^0+e_j^0$ could receive $1$ bits and the nodes $L(C) \setminus (B^0-b_i^0+e_j^0) = E^0 - e_j^0 + b_i^0$ could receive $0$ bits in the following execution of the $AR$ process: the $\gamma=0$ bit is passed down the chunk, flipping at the parent of $b_i^0$ (sending a $0$ bit to $b_i^0$, and a $1$ bit to its sibling), passing the $1$ bit down the chunk, flipping back at the parent of $e_j^0$ (sending a $1$ bit to $e_j^0$, and a $0$ bit to its sibling), and passing the $0$ bit down the rest of the chunk.

The claim follows by a symmetric argument from before.
\end{proof}

\section{Discrepancy Analysis}
\label{sect:analysis}

In this section, we upper bound the discrepancy of our matroid
cups algorithm.

\begin{lemma}\label{lemma:disc_unit}\cite{tijdeman1980,angel2009discrete}
The Unit Clock algorithm has discrepancy at most $1$.
\end{lemma}

\begin{lemma}\label{lemma:disc_chunk} 
Let $D$ be the discrepancy of the Unit Clock algorithm. 
Let $\gamma \in \{0, 1\}$. Let $C$ be a chunk and $w \in L(C)$. Then the stream of bits that the $\gamma$ parity chunk algorithm sends to $w$, with respect to target probability $y_w^{\gamma}$ of $1$ bits, has discrepancy at most $4D$.
\end{lemma}

\begin{proof}
We suppress $\gamma$ from the notation. First, suppose $w = e_i$ for some $e_i \in E$. Consider an interval $(a_{k-1}, a_k)$ which has positive intersection with $(Q_{e, i-1}, Q_{e, i})$. Let $s_k(t)$ be the number of times the chunk clock selects $(a_{k-1}, a_k)$ in its first $t$ calls, and let

\[ \delta_k(t) = s_k(t) - t(a_k-a_{k-1})\]

be the signed discrepancy for interval $(a_{k-1}, a_k)$ with respect to target probability $a_k-a_{k-1}$. If $e_i$ is subordinate in $(a_{k-1}, a_k)$, then let $s_{e_i, k}(m)$ be the number of times the vertex clock $(a_{k-1}, a_k)$ selects $e_i$ in its first $m$ calls, and let

\[ \delta_{e_i, k}(m) = s_{e_i, k}(m) - mp_{e_i, k} \]

be the signed discrepancy for $e_i$ in the vertex clock $(a_{k-1}, a_k)$, with respect to target probability $p_{e_i, k} = \frac{ |  (Q_{e, i-1}, Q_{e, i})  \cap (a_{k-1}, a_{k}) |}{a_k - a_{k-1}}$.

If $e_i$ is dominant in $(a_{k-1}, a_k)$, then after $t$ calls to the chunk algorithm, this interval contributes $s_k(t)$ $1$ bits to $e_i$. If $e_i$ is subordinate in $(a_{k-1}, a_k)$, then after $t$ calls to the chunk algorithm, this interval contributes $s_{e_i, k}(s_k(t))$ bits to $e_i$. Note that

\begin{align*}
s_{e_i, k}(s_k(t)) &= s_k(t)p_{e_i, k} + \delta_{e_i, k}(s_k(t)) \\
&= \left(t(a_k-a_{k-1}) + \delta_k(t)\right)p_{e_i, k} + \delta_{e_i, k}(s_k(t)) \\
&= t(|  (Q_{e, i-1}, Q_{e, i})  \cap (a_{k-1}, a_{k}) |) + p_{e_i, k} \delta_k(t) + \delta_{e_i, k}(s_k(t))
\end{align*}

So the net target probability is $|  (Q_{e, i-1}, Q_{e, i})  \cap (a_{k-1}, a_{k}) |$, and summing over all intervals which have positive intersection with $e_i$, the net target probability for $e_i$ is $|(Q_{e, i-1}, Q_{e, i})| = Q_{e, i} - Q_{e, i-1} = y_{e_i}^{\gamma}$, as desired. Let $s_{e_i}(t)$ be the total number of times that $e_i$ receives a $1$ bit in the first $t$ calls to the chunk algorithm, and let $\delta_{e_i}(t) = s_{e_i}(t) - ty_{e_i}^{\gamma}$.

By maximality of the constructed intervals, an element occurs in at most two consecutive intervals, and if it occurs in two, it is subordinate in the first. Indeed, if an element's interval $(Q_{e, i-1}, Q_{e, i})$ extends beyond the first interval $(a_{k-1}, a_k)$ that it intersects, then the interval $(a_k, a_{k+1})$ will extend to at least the right endpoint $Q_{e, i}$. Further, if the element was dominant in $(a_{k-1}, a_k)$, then that interval already could have extended to the right endpoint $Q_{e, i}$.

This gives four cases, where in each case we use the fact that $|\delta_k(t)| \leq D$ and $|\delta_{e_i, k}(m)| \leq D$, because these are the discrepancies of Unit Clock algorithms.
\begin{enumerate}
\item \(e_i\) is subordinate in one interval \((a_{k-1}, a_k)\). Then
\begin{equation*}
\delta_{e_i}(t)
=p_{e_i,k}\delta_k(t)+\delta_{e_i,k}(s_k(t)),
\end{equation*}
so \(|\delta_{e_i}(t)|\le 2D\).

\item \(e_i\) is dominant in one interval \((a_{k-1}, a_k)\). Then
\begin{equation*}
\delta_{e_i}(t)=\delta_k(t),
\end{equation*}
so \(|\delta_{e_i}(t)|\le D\).

\item \(e_i\) is subordinate in \((a_{k-1}, a_k)\) and dominant in \((a_k, a_{k+1}\). Then
\begin{equation*}
\delta_{e_i}(t)
=p_{e_i,k}\delta_k(t)+\delta_{e_i,k}(s_k(t))
+\delta_{k+1}(t),
\end{equation*}
so \(|\delta_{e_i}(t)|\le 3D\).

\item \(e_i\) is subordinate in both \((a_{k-1}, a_k)\) and \((a_k, a_{k+1})\). Then
\begin{align*}
\delta_{e_i}(t)
={}&p_{e_i,k}\delta_k(t)+\delta_{e_i,k}(s_k(t))+p_{e_i,k+1}\delta_{k+1}(t)
+\delta_{e_i,k+1}(s_{k+1}(t)),
\end{align*}
so \(|\delta_{e_i}(t)|\le 4D\).
\end{enumerate}

The argument for \(b_i \in B\) is symmetric, where we count the number of removals from $B$ with target probability \(1-y_{b_i}\). The ``inclusion'' signed discrepancy is the negative of the ``removal'' signed discrepancy, so its absolute discrepancy is also at most \(4D\).
\end{proof}

\begin{lemma}\label{lemma:disc_fill}
Let $D$ be the discrepancy of the $\gamma$ parity chunk algorithm. Let $u \in U$ be a leaf in the series-parallel forest $T$. Then the Fill algorithm has discrepancy at most $\frac{2D}{1-\rho} + 2D$ with respect to target probability $y_u$.
\end{lemma}

\begin{proof}
Let \(s_v(t)\) count the number of times vertex \(v\) receives bit \(1\) in the first \(t\) steps of the Fill algorithm, and write
\begin{equation*}
\delta_v(t)=s_v(t)-ty_v,
\end{equation*}
where \(y_v\) is its target frequency of receiving bit \(1\).

Consider a chunk \(C\) with root \(v\), and an element \(w\in L(C)\), where \(L(C)\) consists of the children of vertices of \(C\) that lie outside \(C\). For \(\gamma\in\{0,1\}\), let \(s_w^\gamma(m)\) count the number of times \(w\) receives bit \(1\) in the first \(m\) calls to the \(\gamma\) parity chunk algorithm for $C$. Define
\begin{equation*}
\delta_w^\gamma(m)=s_w^\gamma(m)-my_w^\gamma,
\end{equation*}
where \(y_w^\gamma\) is the target frequency of sending bit \(1\) to \(w\) when \(C\) receives bit \(\gamma\). By assumption,
\begin{equation*}
|\delta_w^\gamma(m)|\le D.
\end{equation*}

After \(t\) steps, the \(1\) parity chunk algorithm for \(C\) has been called \(s_v(t)\) times, and the \(0\)-parity chunk algorithm for \(C\) has been called \(t-s_v(t)\) times. Hence
\begin{align*}
s_w(t)
&=s_w^0(t-s_v(t))+s_w^1(s_v(t))\\
&=(t-s_v(t))y_w^0+s_v(t)y_w^1\\
&\qquad+\delta_w^0(t-s_v(t))+\delta_w^1(s_v(t)).
\end{align*}
Since
\begin{equation*}
y_w=(1-y_v)y_w^0+y_vy_w^1,
\end{equation*}
by Lemma \ref{lemma:computey}, subtracting \(ty_w\) gives
\begin{align*}
\delta_w(t)
&=s_w(t)-ty_w\\
&=(t-s_v(t))y_w^0+s_v(t)y_w^1
-t\bigl((1-y_v)y_w^0+y_vy_w^1\bigr)\\
&\qquad+\delta_w^0(t-s_v(t))+\delta_w^1(s_v(t))\\
&=-(s_v(t)-ty_v)y_w^0+(s_v(t)-ty_v)y_w^1\\
&\qquad+\delta_w^0(t-s_v(t))+\delta_w^1(s_v(t))\\
&=(y_w^1-y_w^0)\delta_v(t)
+\delta_w^0(t-s_v(t))+\delta_w^1(s_v(t)).
\end{align*}
Consequently,
\begin{equation*}
|\delta_w(t)|
\le |y_w^1-y_w^0||\delta_v(t)|+2D.
\end{equation*}

First consider the case that $w$ is the root of some chunk in $T$. Via Lemma \ref{lemma:computey}, if $w = w_i$, then we have $y_w^1 - y_w^0 = \alpha(i)$. If $w$ is a child of the last vertex of the chunk $C$, then $y_w^1 - y_w^0 = \alpha(i) \leq \rho$. Otherwise, $y_w^1 - y_w^0 = \alpha(i) \leq \alpha_w \leq 1-\rho$. Thus $y_w^1 -y_w^0 \leq \max\{\rho, 1-\rho\} = \rho$ because $\rho=2/3$. We now show inductively down the chunk tree that every chunk root \(v\) satisfies
\begin{equation*}
|\delta_v(t)|\le \frac{2D}{1-\rho}.
\end{equation*}
At a root of $T$, every incoming bit is \(1\) and \(y_v=1\), so \(\delta_v(t)=0\). For a chunk root \(w\) which is not a root of $T$, the induction hypothesis gives
\[
|\delta_w(t)|
\le \rho|\delta_v(t)|+2D
\le \rho\frac{2D}{1-\rho}+2D
=\frac{2D\rho+2D(1-\rho)}{1-\rho}
=\frac{2D}{1-\rho}.
\]

Finally, let \(u \in U\) be a leaf of $T$, and let \(v\) be the root of the chunk containing its parent. Using \(|y_u^1-y_u^0|\le1\), we obtain
\begin{equation*}
|\delta_u(t)|
\le |\delta_v(t)|+2D
\le \frac{2D}{1-\rho}+2D.
\end{equation*}
$|\delta_u(t)|$ is the discrepancy of the Fill algorithm because the algorithm includes \(u \in U\) precisely when it receives bit \(1\), completing the proof.
\end{proof}

\begin{lemma}\label{lemma:disc_overall}
Let $D$ be the discrepancy of the Fill algorithm. Then overall algorithm has discrepancy at most $D+1$.
\end{lemma}

\begin{proof}
Let $H$ be the block length, and recall that $\ell_u=\lfloor Hx_u\rfloor$ and $y_u=Hx_u-\ell_u$. Let $I_j$ be the fill selected for block $j$, so that this block schedules element $u$ exactly
\begin{equation*}
z_{u,j}=\ell_u+\mathbf{1}[u\in I_j]
\end{equation*}
times. Define the fill signed discrepancy
\begin{equation*}
\delta_u^{\mathrm{fill}}(m)
=\sum_{j=1}^{m}\mathbf{1}[u\in I_j]-my_u,
\end{equation*}
whose absolute value is at most $D$.

Write an arbitrary time as $t=mH+q$, where $0\le q<H$, and let $a_u(q)$ count services of $u$ in the first $q$ steps of block $m+1$. The internal-block guarantee from \cite{Mendoza-CadenaM26} gives
\begin{equation*}
a_u(q)=\frac{q}{H}z_{u,m+1}+\varepsilon_u(q),
\qquad |\varepsilon_u(q)|\le 1.
\end{equation*}
Let $s_u(t)$ count the total services of $u$ by time $t$, and let $\theta=q/H$. Since $Hx_u=\ell_u+y_u$, we have
\begin{align*}
s_u(t)-tx_u
&=m\ell_u+\sum_{j=1}^{m}\mathbf{1}[u\in I_j]
+a_u(q)-(mH+q)x_u\\
&=\sum_{j=1}^{m}\mathbf{1}[u\in I_j]-my_u
+a_u(q)-qx_u\\
&=\delta_u^{\mathrm{fill}}(m)
+\theta z_{u,m+1}+\varepsilon_u(q)
-\theta(\ell_u+y_u)\\
&=\delta_u^{\mathrm{fill}}(m)
+\theta\bigl(\ell_u+\mathbf{1}[u\in I_{m+1}]\bigr)
-\theta(\ell_u+y_u)+\varepsilon_u(q)\\
&=\delta_u^{\mathrm{fill}}(m)
+\theta\bigl(\mathbf{1}[u\in I_{m+1}]-y_u\bigr)
+\varepsilon_u(q)\\
&=\delta_u^{\mathrm{fill}}(m)
+\theta\bigl(\delta_u^{\mathrm{fill}}(m+1)
-\delta_u^{\mathrm{fill}}(m)\bigr)
+\varepsilon_u(q)\\
&=(1-\theta)\delta_u^{\mathrm{fill}}(m)
+\theta\delta_u^{\mathrm{fill}}(m+1)
+\varepsilon_u(q).
\end{align*}
Thus the overall signed discrepancy is a convex combination of the fill signed discrepancies at two consecutive block boundaries, plus the internal-block error. Consequently,
\begin{equation*}
|s_u(t)-tx_u|
\le (1-\theta)D+\theta D+1
=D+1.
\end{equation*}
for all elements $u \in U$ and times $t \geq 0$, completing the proof.
\end{proof}

\begin{lemma}\label{lemma:disc_overall_33}
The overall algorithm has discrepancy at most $33$.
\end{lemma}

\begin{proof}
The Unit Clock algorithm has discrepancy at most $1$ (Lemma \ref{lemma:disc_unit}). Thus the $\gamma$ parity chunk algorithm has discrepancy at most $4$ (Lemma \ref{lemma:disc_chunk}). Thus for $\rho = 2/3$, the Fill algorithm has discrepancy at most $\frac{2(4)}{1-2/3} + 2(4) = 32$ (Lemma \ref{lemma:disc_fill}). Thus the overall algorithm has discrepancy at most $32+1 = 33$ (Lemma \ref{lemma:disc_overall}).
 \end{proof}

\noindent{\bf Acknowledgments:} The result arose from extensive interactions between the authors and GPT-6 Astra. The exposition is primarily the product of the authors. Claude was used to aid the writing of the section on the analysis of discrepancy.  Claude was also used to produce  the figures, create bibtex entries, and create LaTeX code for some of the standard definitions in the appendix. The authors accept full responsibility for the correctness of the result. We made a reasonable attempt to track the source of the ideas, and have included citations as appropriate.

\bibliographystyle{alpha}
\bibliography{bib}

\clearpage
\appendix

\section[Standard Matroid Theory Background]{Standard Matroid Theory Background \cite{schrijver_book}}

\begin{definition}[Matroid]
A \emph{matroid} is a pair $M = (U, \mathcal{I})$, where $U$ is a finite set
(the \emph{ground set}) and $\mathcal{I} \subseteq 2^{U}$ is a family of subsets
of $U$ (the \emph{independent sets}) satisfying:
\begin{enumerate}
  \item[(I1)] $\emptyset \in \mathcal{I}$.
  \item[(I2)] (Hereditary property) If $I \in \mathcal{I}$ and $J \subseteq I$,
        then $J \in \mathcal{I}$.
  \item[(I3)] (Exchange property) If $I, J \in \mathcal{I}$ and $|I| < |J|$,
        then there exists $u \in J \setminus I$ such that
        $I \cup \{e\} \in \mathcal{I}$.
\end{enumerate}
\end{definition}

\begin{definition}[Matroid intersection]
Let $M_1 = (U, \mathcal{I}_1)$ and $M_2 = (U, \mathcal{I}_2)$ be two matroids on
the same ground set $U$, with rank functions $r_1$ and $r_2$. The
\emph{intersection} of $M_1$ and $M_2$ is the set system
\[
  M_1 \cap M_2 = (U, \mathcal{I}_1 \cap \mathcal{I}_2),
\]
whose members are the \emph{common independent sets}: those $I \subseteq U$
that are independent in both $M_1$ and $M_2$. In general $M_1 \cap M_2$ is not a
matroid.

The \emph{matroid intersection problem} asks for a common independent set of
maximum cardinality, and this problem can be solved in polynomial time given efficient independence oracles for $M_1, M_2$.
\end{definition}

\begin{definition}[Matroid independence polytope]
Let $M_1 = (U, \mathcal{I})$ be a matroid. The \emph{matroid independence polytope} is

\[ P(M) = \text{conv}\{\mathbf{1}_I : I \in \mathcal{I} \}\]

where $\mathbf{1}_I$ is the indicator vector of $I \in \mathcal{I}$. It is known that $P(M) = \{x \in [0, 1]^U : x(S) \leq r(S)\}$, where $x(S) = \sum_{u \in S} x_u$ and $r$ is the rank function of the matroid.
\end{definition}

\begin{theorem}
For matroids $M_1, M_2$ on common ground set $U$, the polytope
$P(M_1) \cap P(M_2)$ is integral, that is,
\[
  P(M_1) \cap P(M_2) = \operatorname{conv}\{\,\mathbf{1}_I : I \in \mathcal{I}_1 \cap \mathcal{I}_2\,\}.
\]
\end{theorem}

\clearpage
\section[Proof of Theorem 1]{Proof of Theorem \ref{thm:matroidbamboo}}\label{app:matroidbamboo}

\begin{proof}[Proof of Theorem \ref{thm:matroidbamboo}]
Without loss of generality, assume that $x_u > 0$ for all $u \in U$, because the elements with $x_u = 0$ can simply be removed from $M$ and never scheduled. Let $n = |U|$. We first define modified rates $\hat x$, that round up to the next rate that is an integer multiple of $\frac{1}{k}$,
where $k > 1$ is some parameter which we will specify later. Let $\hat x_u = 1/k$ if $x_u \le 1/k$,
and $\hat x_u =  \lceil x_u k \rceil /k$ otherwise, for all $u \in U$. Then let $\ell = 1+n/k$ and $\tilde x_u = \hat x_u / \ell$.
Note that $\tilde x$ lies inside the independence polytope of $M$, because for all $\emptyset \neq S \subseteq U$,

\[ \tilde{x}(S) = \frac{\hat{x}(S)}{\ell} \leq \frac{x(S)+n/k}{1+n/k} \leq \max\{1, x(S)\} \leq r(S) \]

Our algorithm  then produces the schedule
 in blocks of size $H= k \ell$ time units using the reduction to matroid intersection from \cite{Mendoza-CadenaM26}.
This algorithm guarantees that occurrence $j$ of element
$u \in U$ is in the time range $\left[\left\lfloor \frac{j-1}{\tilde{x}_u}\right\rfloor + 1, \left\lceil \frac{j}{\tilde{x}_u} \right\rceil \right]$ for all $j \in [k\ell \tilde{x}_u]$. Thus

\[ \tilde{h}_u(t) \leq \tilde{x}_u\left(\left(\left\lceil \frac{j+1}{\tilde{x}_u} \right\rceil-1\right) - \left(\left\lfloor\frac{j-1}{\tilde{x}_u}\right\rfloor+1\right)\right) = \tilde{x}_u\left(\left\lceil \frac{j+1}{\tilde{x}_u} \right\rceil - \left\lfloor\frac{j-1}{\tilde{x}_u}\right\rfloor-2\right) \leq \tilde{x}_u\left(\frac{2}{\tilde{x}_u}\right) = 2 \]

where $\tilde{h}_u(t)$ is the height with respect to growth rates $\tilde{x}_u$. Note that for all $\alpha > 0$, if an algorithm  guarantees maximum height at most $h$ for growth vector $y$ then the algorithm guarantees height at most $\alpha h$ for the growth vector $\alpha y$. Because $x \leq \hat{x} = \ell\tilde{x} = (1+n/k)\tilde{x}$, this algorithm guarantees maximum height at most $2(1 + n/k)$ for growth vector $x$. Selecting $k = \lceil 2n/\eps \rceil$ gives that the algorithm guarantees maximum height at most $2+\eps$ and runs in time polynomial in $n, H$, and the input encoding length of $x$, so the algorithm runs in time polynomial in $n$ and $1/\eps$.
\end{proof}

\end{document}